\documentclass[%
reprint,
 amsmath,amssymb,
 aps,
 prb,
]{revtex4-1}
\usepackage{graphicx}% Include figure files
\usepackage{dcolumn}% Align table columns on decimal point
\usepackage{bm}% bold math
\usepackage{hyperref}% add hypertext capabilities
\usepackage[mathlines]{lineno}% Enable numbering of text and display math
\usepackage{color}

\begin{document}

\title{Quantum Confinement and Heavy Surface States of Dirac Fermions in Bismuth (111) Films: an Analytical Approach}% Force line breaks with \\
%\thanks{A footnote to the article title}%

\author{V.V. Enaldiev}
 \email{vova.enaldiev@gmail.com}%
 \affiliation{Kotel'nikov Institute of Radio-engineering and Electronics of Russian Academy of Sciences, 11-7 Mokhovaya St., Moscow, 125009 Russia}%%Lines break automatically or can be forced with \\
\author{V.A. Volkov}%
 \email{volkov.v.a@gmail.com}
 \affiliation{Kotel'nikov Institute of Radio-engineering and Electronics of Russian Academy of Sciences, 11-7 Mokhovaya St., Moscow, 125009 Russia}%
\affiliation{Moscow Institute of Physics and Technology, Institutskiy per. 9, Dolgoprudny, Moscow Region, 141700 Russia}

\date{\today}

\begin{abstract}
Recent high-resolution angle-resolved photoemission spectroscopy experiments have given a reason to believe that pure bismuth is topologically non-trivial semimetal. We derive an analytic theory of surface and size-quantized states of Dirac fermions in Bi(111) films taking into account the new data. The theory relies on a new phenomenological momentum-dependent boundary condition for the effective Dirac equation. The boundary condition is described by two real parameters that are expressed by a linear combination of the Dresselhaus and Rashba interface spin-orbit interaction parameters. In semi-infinite Bi(111), near $\overline{\rm M}$-point the surface states possess anisotropical parabolic dispersion with very heavy effective mass in $\overline{\rm \Gamma}-\overline{\rm M}$ direction order of ten free electron masses, and light effective mass in $\overline{\rm M}-\overline{\rm K}$ direction order of one hundredth of free electron mass. In Bi(111) films with equivalent surfaces, the surface states from top and bottom surfaces are not splitted. In such symmetric film with arbitrary thickness, bottom of the lowest quantum confinement subband in conduction band coincides with the bottom of bulk conduction band in $\overline{\rm M}$-point. 
\end{abstract}

\maketitle

\section{Introduction}

Bismuth takes an important place in solid state physics. High-quality bismuth single crystals were the first electron systems which enabled discovery of fundamental quantum phenomena such as Shubnikov--de Haas \cite{Subnikov_deHaas} and de Haas--van Alphen oscillations \cite{deHaas_vanAlphen} as well as quantum size effect in thin films \cite{Ogrin,sandomir}. Anisotropic dispersion and multi-valley structure of carriers in bismuth attract great attention till nowadays \cite{Feldman,Behnia,Fuseya_Hole_anisotropy,Collaudin,Zhu,Kunchler}. Recently, superconductivity has been observed in pure bismuth under ambient pressure \cite{Prakashaaf}. 

Bismuth-based crystalline compounds was in the first Bi$_{1-x}$Sb$_x$ generation of topological insulators whereas it was believed that pure bismuth is a topologically trivial material \cite{Fu_Kane,Hasan_Kane}, which was derived from pseudopotential \cite{Golin} and tight-binding \cite{Liu_Allen} calculations of bulk band structure. This was also agreed with first-principles calculations of bismuth surface state (SS) spectra \cite{Koroteev_1,Zhang_Bi}, although preceding angle-resolved photoemission spectroscopy (ARPES) results for Bi(111) surface  \cite{Ast_1, Ast_2,Hofmann} can not confirm the theoretical conclusion due to lack of resolution of SS spectra in a vicinity of $\overline{\rm M}$-point of Bi(111) surface Brillouin zone (see Fig.\ref{fig:Brillouin}a). However, recent high resolution ARPES experiments \cite{Ohtsubo,Perfetti,Ito} on Bi(111) surface show that pure bismuth is topologically {\it not trivial} material. Namely, there is a branch in SS spectrum (which was called ''SS1'' in Ref.\cite{Ito}) that starts from valence band at $\overline{\Gamma}$, intersects band gap and ends in a vicinity of conduction band minimum (CBM) in $\overline{{\rm M}}$-point. Existence of this branch is a manifestation of non-trivial topological order in bismuth \cite{Ohtsubo,Perfetti}. There is no contradiction with the topological theory, because relative order of symmetric $L_s$ and antisymmetric $L_a$ bulk bands near Fermi-level plays a key role in topological classification of bismuth. Due to small energy difference between these bands a weak modification of parameters in the Liu-Allen tight-binding model results in inverted band order (i.e. $L_a$ is higher than $L_s$), with little change in rest band structure \cite{Ohtsubo,Perfetti}. Though even in the high-resolution experiments it is not so easy to identify the true dispersion of SSs in vicinity of $\overline{{\rm M}}$-point by virtue of small band gap ($\approx$ 12 meV) and sharp dispersion of bulk bands with respect to spectrum of SSs. Lately SS spectra in Bi(111) thick \cite{Ohtsubo_Kimura} and thin \cite{Saito_Kazuo} films were also studied in frames of tight-binding models and first-principles density functional theory \cite{Kotaka_Ishii}. Authors of Ref.\cite{Ohtsubo_Kimura} suggested a new set of tight-binding parameters in the Liu-Allen model \cite{Liu_Allen} that corresponds to topological non-trivial phase of bismuth and allows one to describe surface spectra around $\overline{{\rm M}}$-point obtained in the high-resolution ARPES experiments \cite{Ohtsubo,Ito}. However, as this model comprises quite a few parameters it is not intuitively clear changes of what parameters make spectra of SSs approach conduction band. Besides, it is not so easy to extend that approach for the case of external electric and magnetic fields. Refs.[\onlinecite{Saito_Kazuo},\onlinecite{Kotaka_Ishii}] consider only SS spectra in ultrathin Bi(111) films, which does not allow one to trace evolution of SSs from bulk limit. %As far as we know the first calculation of Bi(111) SS spectra were carried out in Ref.\cite{Molotkov}, however they predict conical spectra of SS in vicinity of $\overline{{\rm M}}$-point 

In this paper we analytically study spectra of surface and size-quantized states in Bi(111) films in frames of envelope function approximation at the vicinity of $\overline{{\rm M}}$-point, where electrons and holes are described by an effective Dirac equation \cite{Wolff}. Based on derived phenomenological boundary condition (BC) for envelope functions at film's surface and taking into account the new ARPES data, we find that SSs have parabolic dispersion around CBM in vicinity of $\overline{{\rm M}}$-point. We reveal that (i) spectra of SSs for semi-infinite Bi(111) are very anisotropic, (ii) for Bi(111) film SSs from top and bottom surfaces do not couple in case of equivalent surfaces, (iii) there are two degenerate states whose energy equals bulk CBM and does not depend on thickness of the film. We also show evolution of SS spectra in vicinity of $\overline{{\rm M}}$-point as a function of the film thickness. Our theory comprises only two phenomenological parameters whose values may be extracted from ARPES data\cite{Ohtsubo,Perfetti,Benia}. 

The paper is organized as follows. Sec.\ref{section_BC} is devoted to derivation of BC for envelope functions of Dirac fermions in $L$-valley of bismuth at (111) surface taking into account the new ARPES data. In Sec.\ref{section_SSs} we calculate spectra of SSs for semi-infinite sample. Sec.\ref{section_film} is devoted to calculation of spectra for size-quantized and surface states in Bi(111) films. In Sec.\ref{section_summary} we summarize our results. Details of the calculations are given in Appendices \ref{Appendix_A},\ref{Appendix_B},\ref{App3}.

\section{Boundary condition}\label{section_BC}

\begin{figure}
    \centering
    \includegraphics{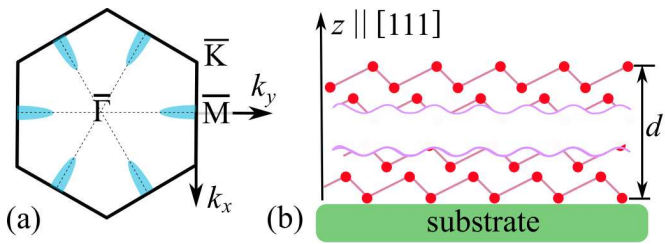}
    \caption{(a) Surface Brillouin zone of Bi(111) film; (b) Schematic side view of Bi(111) film of thickness $d$ with non-equivalent surfaces.}
    \label{fig:Brillouin}
\end{figure}

For simplicity, in the main body of the paper we consider an isotropic Dirac equation  that describes carriers in $L$-valleys of bismuth  \cite{Wolff, Enaldiev} (in Appendix \ref{Appendix_A} we account for bulk anisotropy of bismuth):
\begin{equation}\label{Dirac_eq}
    \left(
        \begin{array}{cc}
        m-E &  v\bm{\sigma p} \\%\sum_{i=x,y,z}v_i\sigma_i p_i \\
        v\bm{\sigma p} & -m-E %\sum_{i=x,y,z}v_i\sigma_i p_i 
        \end{array}
    \right)
    \left(
        \begin{array}{c}
            \Psi_c \\
            \Psi_v
        \end{array}
    \right) = 0,
\end{equation}
where $2m>0$ is the bulk band gap in the $L$-valley, $v$ is absolute value of matrix element for momentum operator between symmetric and anti-symmetric Bloch functions of band extrema, $\bm{\sigma}=\left(\sigma_{x},\sigma_{y},\sigma_{z}\right)$ is vector of the Pauli matrices in standard representation, $\bm{p}$ is three-dimensional quasi-momentum counted from $L$-point of bulk Brillouin zone,  $\Psi_{c,v}$ are spinor envelope functions of states in conductance and valence bands, respectively. Dirac equation (\ref{Dirac_eq}) is written in local Cartesian reference frame shown on Fig.\ref{fig:Brillouin}a,b.

To describe a plane surface in the envelope function approximation neglecting intervalley interaction one should supplement the Dirac equation (\ref{Dirac_eq}) by a BC at the surface. Such a BC couples $\Psi_c$ and $\Psi_v$ \cite{Volkov_Pinsker_English,Enaldiev}: 
\begin{equation}\label{main_BC}
 \Gamma\Psi\equiv 
 \left.\left(
 \begin{array}{cc}
    -M\left(k_{||}\right)  & \sigma_0  \\
      0 & 0 
 \end{array}
 \right) \left(
        \begin{array}{c}
            \Psi_c \\
            \Psi_v
        \end{array}
    \right) \right|_{{\rm surface}} = 0
 %\left[\Psi_v-M\left(k_{||}\right)\Psi_c\right]_{{\rm surface}}=0,
\end{equation}
where the first equality defines BC matrix $\Gamma$ described by 2$\times$2 unit matrix $\sigma_0$ and an unknown matrix $M\left(k_{||}\right)$ that depends, in general, on wave vector components along a surface. General expression for matrix $M$ is fixed by Hermiticity of the Dirac Hamiltonian in confined space \cite{Volkov_Pinsker_English} \mbox{$\bm{\sigma n}M+M^{+}\bm{\sigma n}=0$} and time-reversal symmetry $TM=MT$ described by the operator $T=-i\sigma_yK$. Matrix $M$ satisfying both of these constraints is of the following form (up to the first order in $\bm{k}_{||}$):
\begin{equation}\label{M_gen}
M = ia_0\bm{\sigma n} + i\bm{b}\bm{k}_{||}\sigma_0 + \bm{k}_{||i}\left[C_{ij}-\left(C_{ik}n_k\right)n_j\right]\sigma_j,
\end{equation}
where we imply summation over repeated indexes, $\bm{n}$ is a unit vector normal to a surface, $a_0, \bm{b}, C_{ij}$ are scalar, vector and tensor real phenomenological parameters, that describe properties of a surface. Usually one is only interested in the BC (\ref{main_BC}) with matrix $M$ described by the first term in Eq. (\ref{M_gen}) \cite{Volkov_Pinsker_English,Enaldiev}, as the other terms are small being higher order in wave vector. Such a BC results in conical spectrum of SSs \cite{Volkov_Pinsker_English,Enaldiev} for any values of $a_0$ except $a_0=0$ and $a_0=\infty$. In case of $a_0=0\,\, (a_0=\infty)$ SSs are dispersionless with energy $E_{SS}=m\,\,(E_{SS}=-m)$, i.e. correspond to conduction (valence) band extremum. However, recent experimental ARPES data  \cite{Ito,Ohtsubo,Behnia,Hirahara} reveals  that that SSs possess parabolic dispersion with very heavy effective mass in $\overline{\Gamma}-\overline{{\rm M}}$ direction in vicinity of $\overline{{\rm M}}$-point approaching to the CBM. We argue that such a dispersion may be described by the BC (\ref{main_BC}), (\ref{M_gen}) with $a_0=0$. Effect of small values of $a_0$ on SS spectra is considered in Appendix \ref{Appendix_B}.

At $a_0=0$ matrix $M$ (\ref{M_gen}) still comprises quite a few number of unknown parameters. Their quantity can be reduced by means of spatial symmetry restrictions. Group of $\overline{{\rm M}}$-point of Bi(111) surface Brillouin zone is $C_{2v}$ point group (see Fig.\ref{fig:Brillouin}a). In the reference frame under consideration action of $C_{2}(z)$ rotation and mirror $M_{x,y}$ reflection in plane perpendicular to $x,y$-axis are expressed by matrices $D\left(C_{2}(z)\right)=i\sigma_0\otimes\sigma_z$ and $D(M_{x,y})=i\sigma_{z}\otimes\sigma_{x,y}$ respectively. Invariance of the BC (\ref{main_BC}) under an operation $O=\{C_2(z),M_{x,y}\}$ of $C_{2v}$ point group $D\left(O\right)\Gamma\left(O^{-1}\bm{k}_{||}\right) D^{-1}\left(O\right)=\Gamma$ leads to two parametric form for linear momentum dependent part of the matrix $M$:
\begin{equation}\label{M_final}
    M = a_1\sigma_xk_x + a_2\sigma_y k_y,
\end{equation}
where $a_{1,2}$ are real parameters characterizing surface properties and have length dimension.  In single band limit, the parameters $a_1$, $a_2$ are proportional to Rashba and Dresselhaus interface constants (see Appendix \ref{App3}).   

% Neglecting by this dependence and taking into account time-reversal invariance of the surface the matrix is determined by the only real phenomenological parameter $a_0$: $M=a_0\left[\sigma_x n_x+(v_y/v_x)\sigma_y n_y +(v_z/v_x)\sigma_z n_z\right]$ where $n_{x,y,z}$ are components of unit normal to the surface. The parameter $a_0$ comprises information about properties of surface structure and bulk bands. 

\begin{figure}
    \centering
    \includegraphics[width=0.9\linewidth]{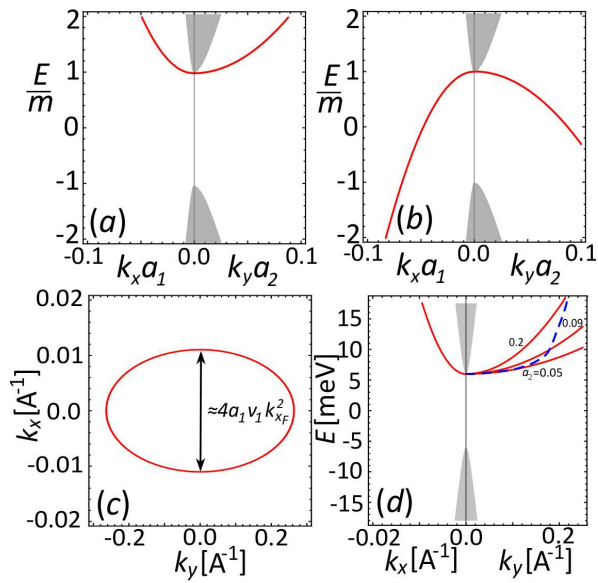}
     \caption{(a,b) Dimensionless SS spectrum (\ref{SS_spectra}) (red) of isotropic Dirac equation with BC (\ref{main_BC}),(\ref{M_final}) along $k_x (k_y)$ in negative (positive) direction at $a_{1,2}>0$ and $a_{1,2}<0$ respectively. (c) Fermi-surface of SSs in semi-infinite Bi(111) with spectra given by Eq.(\ref{SS_spectra_small_moment_App}) with $a_1=10$ A, $a_2=0.2$ A. Value of $a_1$ was extracted from SS Fermi surface in $\overline{{\rm M}}-\overline{{\rm K}}$ direction\cite{Benia} (along $k_x$) at Fermi-level $\approx 19$ meV counted from conduction band minimum. Value of $a_2=0.2$A was derived by comparison with experimental ARPES spectra near the Fermi-level ($\approx 19 meV$) (see subplot (d)). (d) Red curves represent SS spectra in semi-infinite Bi(111) (\ref{SS_spectra_small_moment_App}) along $\overline{{\rm M}}-\overline{{\rm K}}$ direction (negative axis) and $\overline{\Gamma} - \overline{{\rm M}}$ direction (positive axis). In $\overline{{\rm M}}-\overline{{\rm K}}$ direction, the SS spectra is characterized by the parameter $a_1=10$ A that was derived from Fermi-surface size in that direction on the subplot (c). In $\overline{\Gamma} - \overline{{\rm M}}$ direction we plot SS spectra for three different values of the parameter $a_2=0.05,0.09,0.2$A  fitting in the best way dashed curve at different momenta. The dashed curve represents dispersion of SSs extracted from ARPES data of Ref.[\onlinecite{Ohtsubo}]. Bulk parameters of Bi are the following: $2m=12$ meV, $v_1=1.2\times 10^6$ m/s, $v_2=0.1\times 10^6$ m/s, $v_3=1.0\times 10^6$ m/s. Values of velocity matrix elements were derived from the bulk effective mass tensor calculated in Ref.[\onlinecite{Liu_Allen}]. Grey color fills projection of bulk state spectrum.}
    \label{fig:SS}
\end{figure}

\section{Surface states spectra of Dirac fermions}\label{section_SSs}

Here, we consider SSs in a semi-infinite problem for the isotropic Dirac equation (\ref{Dirac_eq}) with the BC (\ref{main_BC}),(\ref{M_final}). Wave functions of SSs in a crystal filling semi-space $z>0$ %with (111) surface orientation, 
are of the form $\phi_{k_{||}}e^{-\kappa z + i{\bf k}_{||}{\bf r}_{||}}$, where $\phi_{k_{||}}$ is a bi-spinor obeying the Dirac equation as well as the BC, $\kappa=\left(m^2+\hbar^2v^2k_{||}^2-E^2\right)^{1/2}/\hbar v>0$ is an inverse decay length of SSs. Spectra of the SS have the following form (with $\hbar=1$):
\begin{equation}\label{SS_spectra}
E_{SS}\left(k_x,k_y\right) = \dfrac{2v \left(a_1 k_x^2 + a_2 k_y^2\right)}{1+ a_1^2 k_x^2 + a_2^2 k_y^2} + m\dfrac{1 - a_1^2 k_x^2 - a_2^2 k_y^2}{1 + a_1^2 k_x^2 + a_2^2 k_y^2}.
\end{equation}
%The SS spectra (\ref{SS_spectra}) is highly anisotropic (see Fig.\ref{fig:SS}a), which is mostly determined by the BC (\ref{main_BC}),(\ref{M1_matrix}) rather than anisotropy of the bulk. 
Spectrum (\ref{SS_spectra}) is shown on Fig.\ref{fig:SS}(a,b) in dimensionless axes. For small wave vectors the SS dispersion is reduced to \mbox{$E_{SS}\approx m + 2va_1k_x^2(1-ma_1/v) + 2va_2k_y^2(1-ma_2/v)$}. Due to smallness of the band gap in real bismuth, curvature of the SS spectrum around $\overline{{\rm M}}$-point is only determined by the parameter $a_{1,2}$ (i.e. $1-ma_{1,2}/v\approx 1$) along $x,y$-axis. In opposite limit $|k_{x,y}a_{1,2}|\gg 1$ SS spectrum saturates to $E_{SS}\approx \left(2v/a_{1,2}\right)(1-ma_{1,2}/2v)$ in $k_x,k_y$-direction. Generally, the SS spectrum is anisotropic even for the isotropic Dirac equation in the bulk. In Appendix \ref{Appendix_A} we calculate SS spectrum taking into account bulk anisotropy of bismuth and extract value of the boundary parameters from Bi(111) ARPES data of Refs.[\onlinecite{Ohtsubo},\onlinecite{Benia}] (see Fig.\ref{fig:SS}c,d): $a_1=10$ A, $0.05{\rm A}\leq a_2\leq 0.2 {\rm A}$. So, effective masses of SSs in semi-infinite Bi(111) (\ref{SS_spectra_small_moment_App}) are equal to \mbox{$m_{xx}=\hbar/4a_1v_1=0.024m_0$} along the $\overline{\rm M}-\overline{\rm K}$ direction and \mbox{$m_{yy}=\hbar/4a_2v_2$} in the range $14m_0\leq m_{yy}\leq 58m_0$ along the $\overline{\rm \Gamma}-\overline{\rm M}$ direction ($m_0$ is the free electron mass).

\section{Spectra of Dirac fermions in films}\label{section_film}

In this section we consider a film of thickness $d$. In general case, two film surfaces are non-equivalent ones (f.e. top is exposed to vacuum but bottom contacts with substrate) and characterized by the BC (\ref{main_BC}), (\ref{M_final}) with the different boundary parameters $a_{1t,b},a_{2t,b}$ for top ($z=d/2$) and bottom ($z=-d/2$) surfaces, respectively. Wave functions of states in the film are of the form $\left[\phi^{(1)}_{k_{||}}e^{-\kappa(z+d/2)} + \phi^{(2)}_{k_{||}}e^{\kappa(z-d/2)}\right] e^{i{\bf k}_{||}{\bf r}_{||}}$. The wave functions with real positive $\kappa$ describe SSs, and with purely imaginary $\kappa$ -- size-quantized states. After substitution the latter wave function to the BC we obtain dispersion equation:
\begin{align}\label{gen_disp}
\left(E - E_{SS}^t\right)\left(E - E_{SS}^b\right)\sin^2\left(k_zd\right) - \qquad\qquad\qquad\qquad\qquad \nonumber\\ \dfrac{v^2k_z^2\left[\left(a_{1b}-a_{1t}\right)^2k_x^2 + \left(a_{2b}-a_{2t}\right)^2k_y^2\right]}{\left[1+a_{1t}^2k_x^2+a_{2t}^2k_y^2\right]\left[1+a_{1b}^2k_x^2+a_{2b}^2k_y^2\right]}=0,
\end{align}
where $k_z=i\kappa$, $E_{SS}^{t/b}$ are spectra of SSs (\ref{SS_spectra}) at top/bottom surfaces in the limit $d\to\infty$. 

\begin{figure}
    \centering
    \includegraphics[width=1.0\linewidth]{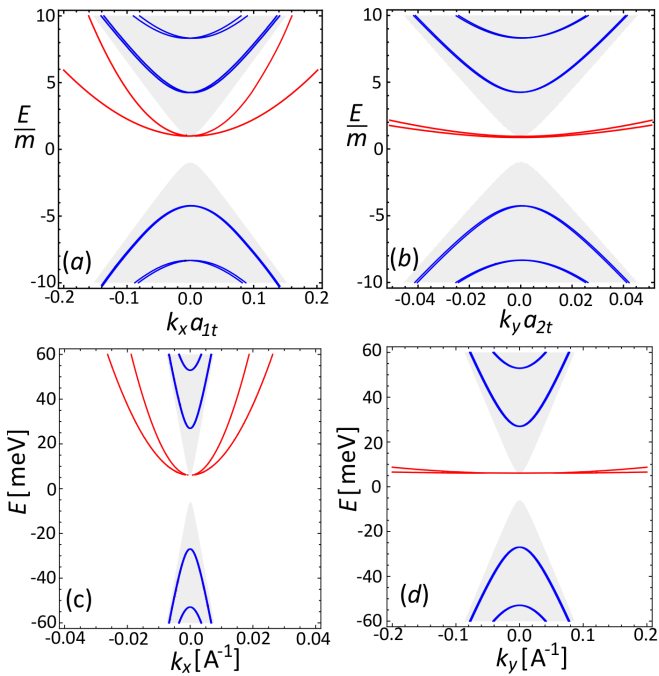}
    \caption{(a,b) Dimensionless SS (red) and size-quantized (blue) spectra of isotropic Dirac equation (\ref{Dirac_eq}) with BC (\ref{main_BC}),(\ref{M_final}) along $k_x,k_y$ directions correspondingly in a film with non-equivalent surfaces described by dimensionless parameters $md/\hbar v=0.76$, $\hbar v/ma_{1t}=219$, $\hbar v/ma_{2t}=65$, $a_{1b}/a_{1t}=0.67$, $a_{2b}/a_{2t}=9.33$. (c,d) Spectra of states in Bi(111) film of 78.366 nm thickness (200 bilayers) in vicinity of $\overline{{\rm M}}$-point. Bulk bismuth parameters are the same as on Fig.\ref{fig:SS}d, boundary parameters are the following ones $a_{1t}=10 {\rm A}, a_{2t}=0.05{\rm A}, a_{1b}=5{\rm A}, a_{2b}=0.01{\rm A}$. For real bismuth parameters size-quantized bands are very weakly splitted even for non-equivalent surfaces. Lightgrey color fills projection of  bulk bands  in vicinity of  $\overline{{\rm M}}$-point.  }
    \label{fig:Film_spectra}
\end{figure}

\begin{figure}
    \centering
    \includegraphics{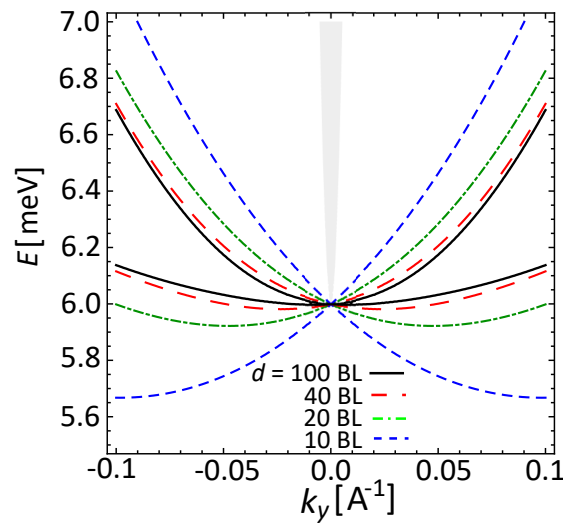}
    \caption{\label{Fig4} SS spectra of Bi(111) film given by Eq.(\ref{gen_disp_app}) with different thickness around conduction band minimum in vicinity of $\overline{{\rm M}}$-point: blue dashed curve $d=3.696$ nm (10 bilayers), green dashed-dotted curve $d=7.626$ nm (20 bilayers), red long dashed curve $d=15.486$ nm (40 bilayers), black thick solid curve $d=39.066$ nm (100 bilayers). Light-grey color fills projection of conduction band. Figure is plotted at the same parameters as on Fig.\ref{fig:Film_spectra}(c,d).  }
    \label{fig:Fermi_arcs}
\end{figure}

In the case of equivalent film surfaces (i.e. $a_{1t}=a_{1b}$, $a_{2t}=a_{2b}$) spectra of size-quantized subbands are determined by the equation $k_zd=\pi n$, $n=1,2,\dots$ All subbands are double degenerate on spin quantum number. A zero band $n=0$ has specific character, two degenerate states with $k_z=k_x=k_y=0$ emerge at CBM with energy independent on the film thickness. These two states possess coordinate independent wave functions $\psi_{\uparrow}\propto (1,0,0,0)^T$, $\psi_{\downarrow}\propto (0,1,0,0)^T$. %The presence of such two states in the bulk CBM looks like violation of the uncertainty principle. 
It may explain very weak thickness dependence of ARPES response from conduction band extrema in Ref.[\onlinecite{Ito}]. With increase of momenta, states in the zero subband become more decaying inside the film finally turning to SSs. %As a function of longitudinal quasi-momentum these states are continuously connected to SSs.   

In the opposite case of non-equivalent surfaces, size-quantized subbands are splitted in spin quantum number (see Fig.\ref{fig:Film_spectra}). We mention that the two degenerate states with energy at bulk CBM also exist in this case.  

As it follows from dispersion equation (\ref{gen_disp}) SSs from top and bottom surfaces do not interact to each other in case of equivalent surfaces. Their spectra are determined by Eq.(\ref{SS_spectra}) and have double degeneracy. Coupling of SSs from top and bottom surfaces emerges in degree of non-equivalence of two surfaces. This interaction leads to anisotropic repulsion of SS spectra (see Fig.\ref{fig:Film_spectra}). There are no such effects for Dirac fermions with momentum-independent BC \cite{Volkov_Pinsker_English} as well as for open BC in the topological insulator films \cite{Linder,Shan}.

On Fig.\ref{Fig4} we demonstrate dependence of SS spectra at energies around CBM on thickness of the Bi(111) film. The less thickness of the film the more spin splitting of SS spectra. However, we stress that thinning of the film does not affect the double degeneracy at $k_x=k_y=0$.  

\section{Summary}\label{section_summary}

In conclusion, we derived a phenomenological momentum dependent BC for the Dirac equation that allow us to quantitatively describe SS spectra in vicinity of $\overline{{\rm M}}$-point in Bi(111) films. The BC is characterized by two real phenomenological parameters that determine effective mass of the SSs in the $\overline{\Gamma}-\overline{{\rm M}}$ and $\overline{{\rm M}}-\overline{{\rm K}}$ directions respectively. The phenomenological parameters are concerned with the parameters of interface Rashba and Dresselhaus spin-orbit interactions. We extract values of the parameters from comparison with recent ARPES data \cite{Ito,Benia} for Bi(111) samples. In the $\overline{\Gamma}-\overline{{\rm M}}$ direction the SSs possess very heavy effective mass in range of $14m_0\leq m_{yy}\leq 58m_0$ depending on energy, but in the $\overline{{\rm M}}-\overline{{\rm K}}$ direction the SSs have light effective mass $m_{xx}=0.024m_0$. The above BC also leads to unusual interaction of SSs from top and bottom surface of the film. Namely, SSs from two film surfaces weakly interact to each other and only in degree of non-equivalence of the surfaces. In addition, our BC results in existence of two degenerate states in Bi(111) films with energy at bulk conduction band minimum regardless of the film thickness. In fact, this energy equals that of bottom of the zeroth subband (\mbox{$n=0$}). Finally, it should be mentioned that the developed approach allows straightforward physically clear generalisation, unlike {\it ab initio} calculations, for the case of smooth external fields. For example, our results can be used to construct a theory for conductivity of the Bi(111) films in which surface state contribution coexists with that of size-quantized states, like in very recent expriments\cite{kroger}.  

We acknowledge support by the Russian Science Foundation (project no. 16-12-10411). 

\appendix
\section{Surface States dispersion around $\overline{{\rm M}}$-point in Bi(111)} \label{Appendix_A}

\begin{figure}
    \centering
    \includegraphics{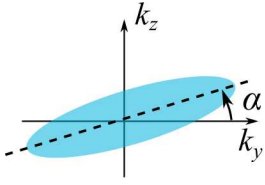}
    \caption{Sketch of projection of bulk electron Fermi-surface in $L$-valley of Bi onto mirror plane ($k_z$-axis is along [111] direction, $k_y$-axis along bisectrix axes, angle $\alpha\approx 6^{\circ}$ \cite{Liu_Allen}).}
    \label{fig:1}
\end{figure}

In $L$-valley of bulk bismuth carriers possess anisotropic pseudo-relativistic dispersion and are described by effective anisotropic Dirac equation:
\begin{equation}\label{Dirac_eq_app}
    \left(
        \begin{array}{cc}
        m-E &  H_{21} \\%\sum_{i=x,y,z}v_i\sigma_i p_i \\
        H_{21}^{+} & -m-E %\sum_{i=x,y,z}v_i\sigma_i p_i 
        \end{array}
    \right)
    \left(
        \begin{array}{c}
            \Psi_c \\
            \Psi_v
        \end{array}
    \right) = 0,
\end{equation}
where 
\begin{align}
 H_{21}=v_1\sigma_xp_x + v_2\sigma_y\left(p_y\cos\alpha-p_z\sin\alpha\right) +\nonumber \\  v_3\sigma_z\left(p_y\sin\alpha+p_z\cos\alpha\right),\nonumber
\end{align}
$2m$ is the bulk band gap in $L$-valley,  $v_{1,2,3}$ are absolute values of matrix elements for respective components of momentum operator between symmetric and anti-symmetric Bloch functions of band extrema, $\sigma_{x,y,z}$ are the Pauli matrices in standard representation, $p_{x,y,z}$ are components of quasi-momentum counted from $L$-point of bulk Brillouin zone, $\alpha$ is an angle ($\approx 6^{\circ}$) between long axes of bulk electron ellipsoidal Fermi surface and $y$-axis (see Fig.\ref{fig:1}). Dirac equation (\ref{Dirac_eq_app}) is written in reference frame where $z$-axis along $[111]$ direction, $x$-axis is perpendicular to mirror plane. 

To derive surface state dispersion we look for general boundary condition (BC) for the Dirac equation in the form used in the main text:
\begin{equation}\label{main_BC_app}
\left.\Gamma\Psi\right|_{{\rm surface}}\equiv
\left(\begin{array}{cc}
    -M\left(k_{||}\right) & 1 \\
    0 & 0 
\end{array}
\right)
\left(
\begin{array}{c}
     \Psi_c  \\
      \Psi_v
\end{array}
\right)
= 0
\end{equation}
Following the main text we use Hermiticity of the Dirac Hamiltonian 
\begin{equation}
\left[\Psi_{c}^{+}\frac{\partial H_{21}}{\partial\bm{p}}\bm{n}\Psi_v + \Psi_{v}^{+}\frac{\partial H_{21}^{+}}{\partial\bm{p}}\bm{n}\Psi_c\right]_{{\rm surface}}=0,
\end{equation}
and time-reversal symmetry $T\Gamma=\Gamma T$ to find general expression for the matrix $M$ up to the linear terms in momentum parallel to the surface:  
\begin{multline}\label{M_gen_app}
    M = ia_0\left[n'_x\sigma_x + \left(v_2/v_1\right)n'_y\sigma_y + \left(v_3/v_1\right)n'_z\sigma_z\right] + \\
    i\bm{b}\bm{k}_{||}\sigma_0 + k_{||i}\left[C_{ij}n'_ln'_l - \left(C_{ik}n'_{k}\right)n'_j\right]\sigma_j
\end{multline}
here $\bm{n}'=\left(n_x,n_y\cos\alpha-n_z\sin\alpha,n_z\cos\alpha+n_y\sin\alpha\right)$, $\bm{n}=(n_x,n_y,n_z)$ is unit normal to a surface, $a_0$, $\bm{b}$, $C_{ij}$ is real scalar, vector and tensor parameters characterizing surface properties. Together with invariance of the BC (\ref{main_BC_app}) under any operation $O$ of $C_{2v}$ point group $D\left(O\right)\Gamma\left(O^{-1}\bm{k}_{||},O^{-1}\bm{n}'\right) D^{-1}\left(O\right) = \Gamma$, we obtain the following form of $M$ matrix:
\begin{multline}\label{M_fin_app}
     M = i a_0\left[-\left(v_2/v_1\right)\sigma_y\sin\alpha + \left(v_3/v_1\right)\sigma_z\cos\alpha\right] + \\
    a_1\sigma_x k_x + a_2\sigma_y k_y\cos\alpha + a_2(v_2/v_3)\sigma_z k_y\sin\alpha
\end{multline}

%It is known \cite{Enaldiev,Volkov_Pinsker_English} that the Dirac equation with the BC (\ref{main_BC_app}) described by only the first term in previous formula leads to conical spectra of SSs for any values of $a_0$ except $a_0=0$ and $a_0=\infty$. In case of $a_0=0\,\, (\infty)$ SSs are dispersionless with energy $E_{SS}=m\,\,(-m)$ (i.e. correspond to conduction (valency) band extremum for all momenta). Experimental data for Bi(111) surface \cite{Ohtsubo,Ito} reveals that SSs possess very heavy effective mass around $\overline{{\rm M}}$-point of surface BZ in $\overline{\Gamma}-\overline{{\rm M}}$ direction. We argue that such a dispersion may be described by the BC (\ref{main_BC_app}), (\ref{M_fin_app}) with $a_0=0$. Effect of small values of $a_0$ on SS spectra will be discussed below.
Below, we first derive SS spectra for anisotropic Dirac equation (\ref{Dirac_eq_app}) with the BC (\ref{main_BC_app}), (\ref{M_fin_app}) at $a_0=0$, and then consider effect of small $a_0$ ($a_0\ll 1$) on spectra and density of SSs in isotropical limit $v_1=v_2=v_3=v$.

Let us now find dispersion of SSs in $L$-valley of semi-infinite Bi(111) crystal filling semi-space $z>0$. SS wave functions are of the form $\phi_{k_{||}}e^{-\left(\kappa'+i\kappa'' \right)z + i{\bf k}_{||}{\bf r}_{||}}$, where $\phi_{k_{||}}$ is a bi-spinor obeying the Dirac equation (\ref{Dirac_eq_app}) as well as the BC (\ref{main_BC_app}), (\ref{M_fin_app}) with $a_0=0$, $\kappa'=\sqrt{\left(m^2+v_1^2k_x^2-E^2\right)/\widetilde{v}^2+v_3^2v_2^2k_y^2/\widetilde{v}^4}>0$ is an inverse decay length of SSs, $\kappa''=k_y\left(v_3^2-v_2^2\right)\sin\alpha\cos\alpha/\widetilde{v}^2$ and $\widetilde{v}^2=v_3^2\cos^2\alpha+v_2^2\sin^2\alpha$. After some algebra one finds expression for SS spectra (with $\hbar=1$):
\begin{equation}\label{SS_spectra_app}
E_{SS}\left(k_x,k_y\right) = \dfrac{2 \left(a_1 v_1 k_x^2 + a_2 v_2 k_y^2\right)}{1+\left(a_1^2 k_x^2 + a_2^2 k_y^2\right)} + m\dfrac{1 - \left(a_1^2 k_x^2 + a_2^2 k_y^2\right)}{1 + \left(a_1^2 k_x^2 + a_2^2 k_y^2\right)}.
\end{equation}
At small momenta the previous equation is reduced to the following:
\begin{equation}\label{SS_spectra_small_moment_App}
\begin{split}
E_{SS}\left(k_x,k_y\right)\approx m + 2a_1v_1\left(1 - ma_1/v_1\right)k_x^2 + 2a_2v_2\left(1 - ma_2/v_2\right)k_y^2  \\
 \approx m + 2a_1v_1k_x^2 + 2a_2v_2k_y^2,\qquad\qquad\qquad\qquad\qquad\qquad
\end{split}
\end{equation}
where the last equality is valid for bismuth due to small band gap. Eq.(\ref{SS_spectra_small_moment_App}) was used in plotting Fermi-surface of SS on Fig.\ref{fig:SS}c.

In case of Bi(111) film of thickness $d$ wave functions of states in the film are of the form $\left[\phi^{(1)}_{k_{||}}e^{-\kappa_1(z+d/2)} + \phi^{(2)}_{k_{||}}e^{-\kappa_2(z-d/2)}\right]e^{{\bf k}_{||}{\bf r}_{||}}$, where $\kappa_{1,2}=\pm\kappa'+i\kappa''$ are two roots of characteristic equation. Wave functions with real positive $\kappa'$ describe SSs, and with purely imaginary $\kappa'$ -- size-quantized states.  After substitution the latter wave function to the BC we obtain dispersion equation:
\begin{align}\label{gen_disp_app}
\left(E - E_{SS}^t\right)\left(E - E_{SS}^b\right)\sin^2\left(k_zd\right) - \qquad\qquad\qquad\qquad\qquad \nonumber\\ \dfrac{\widetilde{v}^2k_z^2\left[\left(a_{1b}-a_{1t}\right)^2k_x^2 + \left(a_{2b}-a_{2t}\right)^2k_y^2\right]}{\left[1+a_{1t}^2k_x^2+a_{2t}^2k_y^2\right]\left[1+a_{1b}^2k_x^2+a_{2b}^2k_y^2\right]}=0,
\end{align}
where $k_z=i\kappa'$, $E_{SS}^{t,b}$ is SS dispersion (\ref{SS_spectra}) with $a_{1t,b}$, $a_{2t,b}$ at top/bottom surface. In case of equivalent surfaces $a_{1t}=a_{1b}$, $a_{2t}=a_{2b}$ dispersion equation (\ref{gen_disp_app}) is reduced to the following one:
\begin{equation}
\left(E - E_{SS}^t\right)\left(E - E_{SS}^b\right)\sin^2\left(k_zd\right)=0.
\end{equation}

\section{Surface state dispersion with account of $a_0$}\label{Appendix_B}
\begin{figure}
    \centering
    \includegraphics{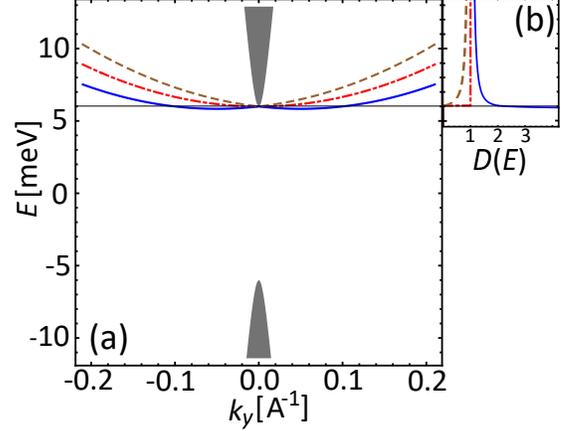}
    \caption{ (a) Red dashed-dotted curve shows SS spectra (\ref{SS_spectra_app}), brown dashed and blue solid curves represent SS spectra (\ref{SS_a_0_app}) with $a_0=-0.005$ and $a_0=0.005$, respectively. Other parameters are the following: $2m=12$meV, $\hbar v=0.65$eV$\cdot$A ($v_1=v_2=v$), $a_1=a_2=0.05$A.  Grey color shows projection of bulk bands. (b) Density of SS with corresponding spectra on Fig.6(a) in units of: $1/8\pi\hbar v a_1$.}
    \label{fig:Compare}
\end{figure}

Here we consider effect of small $a_0$ on SS spectra. For simplicity we neglect by anisotropy $v_1=v_2=v_3=v$ and suppose that $\alpha=0$. Then dispersion equation for SSs reads as follows
\begin{multline}
    m\left(1-a_1^2 k_x^2-a_2^2k_y^2+a_0^2\right) - E\left(1 + a_1^2 k_x^2 + a_2^2k_y^2 - a_0^2\right) + \\
    2v\left(a_1k_x^2+a_2k_y^2\right) =  2a_0\sqrt{m^2+v^2\left(k_x^2+k_y^2\right) - E^2}.
\end{multline}
In the most interesting case $|a_0|\gg|a_{1,2}k_{x,y}|$ we obtain the following SS spectra:
\begin{equation}\label{SS_a_0_app}
\begin{split}
E_{SS} = s\dfrac{2a_0v}{1+a_0^2}\sqrt{k_x^2+k_y^2} + m\frac{1-a_0^2}{1+a_0^2} + \qquad\qquad\qquad \\ 2v\left(a_1k_x^2+a_2k_y^2\right)\dfrac{1-a_0^2}{\left(1+a_0^2\right)^2}.
\end{split}
\end{equation}
In case of $a_0>0$, $s=+ 1$ is valid only for $k_{||}<2a_0mv/\hbar|1-a_0^2|$, $s=-1$ holds for every wave vector. In case of $a_0<0$ only $s=-1$ is allowed for all wave vectors. For small enough momenta in the previous formula the first two term are leading ones and result in conical spectra \cite{Enaldiev,Volkov_Pinsker_English}. The last term gives parabolic curvature and is essential for high enough momenta and $|a_0|\ll 1$.  We compare SS spectra   (\ref{SS_spectra_app}) with spectra (\ref{SS_a_0_app}) at $|a_0|\ll 1$ on Fig.\ref{fig:Compare}.

\section{Relation of parameters $a_{1,2}$ with Rashba and Dresselhaus interface parameters }\label{App3}

In this section we show relation between the boundary parameters $a_1$, $a_2$ in Eq.(\ref{M_final}) and parameters of interface Rashba and Dresselhaus spin-orbit interactions in a single band limit (i.e. for DFs with energies $|E-m|\ll m$). In Ref.[\onlinecite{Devizorova}] values of the Rashba and Dresselhaus interface parameters for GaAs/AlGaAs interface were determined by comparison with experiments on electron spin resonance. Expressing  $\Psi_{v}\approx v\bm{\sigma}\bm{p}\Psi_{c}/2m $ with the help of Eq.(\ref{Dirac_eq}) and multiplying Eq.(\ref{main_BC}) from the left by $-i\sigma_z$ we obtain the following BC:
\begin{equation}
\left.\left[-i\hat{p}_z + \left(1-\frac{2ma_1}{v\hbar}\right)\sigma_yp_x - \left(1-\frac{2ma_2}{v\hbar}\right)\sigma_xp_y\right]\Psi_{c}\right|_{z=0}=0.
\end{equation}  
After $\pi/4$-rotation around $z$ axis and unitary transformation $\widetilde{\Psi}_c=U\Psi_c$ with $U={\rm diag}\left(1, e^{-i\frac{3\pi}{4}}\right)$, the previous equation reads as follows:
\begin{equation}
\left.\left[-i\hat{p}_z + \chi\left(\sigma_xp_y - \sigma_yp_x\right) +  \gamma\left(\sigma_xp_x-\sigma_yp_y\right)\right]\widetilde{\Psi}_{c}\right|_{z=0}=0,
\end{equation}
where parameters $\chi$ and $\gamma$ are determined by the following equalities:
\begin{equation}\label{RD_param}
\begin{split}
\chi = 1 - \frac{m}{\hbar v}\frac{a_1+a_2}{2} \\
\gamma = \frac{m}{\hbar v}\left( a_2 - a_1\right )
\end{split}
\end{equation}
The parameter $\chi$ describes interface Rashba interaction, but $\gamma$ accounts for Dresselhaus interface interaction\cite{Devizorova}. Solving (\ref{RD_param}) with respect to $a_{1,2}$ we obtain the final expression for them via $\chi$ and $\gamma$:
\begin{equation}
\begin{split}
a_1 = \frac{\hbar v}{m}\left[1 - \chi -\gamma/2 \right] \\
a_2 = \frac{\hbar v}{m}\left[1 - \chi +\gamma/2 \right]
\end{split}
\end{equation}

\bibliography{Bibgraph}

\end{document}